\documentstyle[11pt,epsfig]{article}
\def\G{GX~301--2}   \def\B{{\em BeppoSAX}}
\def\mcc#1{\multicolumn{1}{c}{#1}}
\def\Iave{\mbox{$\langle I \rangle$}}
\def\sec#1{\vskip 1\baselineskip {\noindent\large\bf #1} \smallskip}
\def\subsec#1{\vskip 1\baselineskip {\noindent\large\bf #1} \smallskip}

\makeatletter
\def\@cite#1#2{(#1\if@tempswa , #2\fi)}
\def\@biblabel#1{}
\newlength{\bibhang}
\def\bibtitle{REFERENCES}
\def\bibheadtitle{REFERENCES}

\def\thebibliography#1{\@mkboth{\bibheadtitle}{\bibheadtitle}
  \addcontentsline{toc}{section}{\protect\numberline{}\bibtitle}\small\list
  {\relax}{\setlength{\labelsep}{0em}
        \setlength{\labelwidth}{0em}
        \setlength{\itemsep}{0em}
	\setlength{\itemindent}{-\bibhang}
	\setlength{\leftmargin}{\bibhang}}
    \def\newblock{\hskip .11em plus .33em minus .07em}
    \sloppy\clubpenalty4000\widowpenalty4000
    \sfcode`\.=1000\relax}

\def\@citex[#1]#2{\if@filesw\immediate\write\@auxout{\string\citation{#2}}\fi
  \def\@citea{}\@cite{\@for\@citeb:=#2\do
    {\@citea\def\@citea{; }\@ifundefined
       {b@\@citeb}{{\bf ?}\@warning
       {Citation `\@citeb' on page \thepage \space undefined}}%
{\csname b@\@citeb\endcsname}}}{#1}}
\makeatother

\begin{document}

\strut\bigskip\bigskip

\noindent\begin{flushleft}
{\Large\bf BeppoSAX OBSERVATIONS OF AN ORBITAL CYCLE OF \\
       THE X--RAY BINARY PULSAR \G} \\[1.7\baselineskip]
{\large M.~Orlandini$^1$, D.~Dal~Fiume$^1$, F.~Frontera$^{1,2}$,
T.~Oosterbroek$^3$, A.N.~Parmar$^3$,\\
A.~Santangelo$^4$, A.~Segreto$^4$} \\[1.7\baselineskip]
{\em $^1$ TeSRE/C.N.R., via Gobetti 101, 40129 Bologna, Italy \\
 $^2$ Physics Dept.\ Ferrara University, via Paradiso 12, 44100 Ferrara, Italy
\\
 $^3$ SSD/ESA, ESTEC, Keplerlaan 1, 2200 AG Noordwijk, The Netherlands \\
 $^4$ IFCAI/C.N.R., via La Malfa 153, 90146 Palermo, Italy} \\[2.5\baselineskip]
\end{flushleft}

\noindent {\large\bf ABSTRACT}

\smallskip

\noindent We present preliminary results on our campaign of observations of the
X--ray binary pulsar GX301--2. \B\ observed this source six times in
January/February 1998: at the periastron and apoastron, and at other four,
intermediate, orbital phases. We present preliminary results on the \G\
spectral and temporal behaviour as a function of orbital phase.

\sec{1\quad INTRODUCTION}

\noindent The X--ray binary pulsar \G\ (4U~1223--62) is a $\sim 700$~s pulsator
orbiting the B2 Iae supergiant Wray 977 every 41.5 days along the most
eccentric orbit among X--ray binary pulsars \cite{816,1565}. It exhibits a
flaring activity that shows its maximum $\sim 1.4$ days before the periastron
passage. This anticipation with respect to the phase of closest approach to
Wray 977 has been explained as due to the crossing of the neutron star through
a circumstellar disk around the supergiant \cite{406}. Recent
observations by BATSE \cite{406} have revealed the presence of a second flare
occurring near the apoastron, at orbital phase 0.45.

\smallskip

The overall power-law X--ray spectrum of \G\ is characterized by a strong soft
excess below 4 keV. Because pulsation was not detected in this band, the
partial covering model has been ruled out as possible origin of this excess.
ASCA observations revealed the presence of two soft components: a {\em
scattering} component, between 2 and 4 keV, due to the gas stream from the
supergiant to the neutron star, and a {\em ultrasoft} component, below 2 keV,
described with thermal emission from a plasma at $\sim 0.8$~keV \cite{405}. A
strong narrow iron emission line and an absorption edge are also present. The
emission line is the result of fluorescence in the cooler circumstellar
material, while the absorption edge is due to the same material when it crosses
the line of sight and absorbs the X--rays coming from the neutron star. At high
(E$>$10~keV) energies, the power law spectrum is modified by a high energy
cutoff. A possible cyclotron resonance feature at $\sim 40$ keV has been
claimed \cite{407}.

\sec{2\quad OBSERVATIONS}

\noindent \G\ was the target of a campaign of observations with the
Italian-Dutch satellite \B\ \cite{1530}. The source was observed at six
different orbital phases (see Table~\ref{log}), in order to monitor its
spectral and timing behaviour along the orbit. All four Narrow Field
Instruments aboard \B\ worked nominally during all the observations, namely the
LECS (0.1--10 keV), MECS (1.5--10 keV), HPGSPC (3--120 keV), and PDS (15--200
keV).

\smallskip

The source displayed its maximum intensity not at the periastron observation
OP3428/9 but at orbital phase 0.85 (OP3373). The net count rate in this
observation was about four times that detected at the periastron. We confirmed
the increase in the intensity near the apoastron.

\newlength{\lenone}   \settowidth{\lenone}{3514}
\newlength{\lentwo}   \settowidth{\lentwo}{30/1/98 07:19:12}
\newlength{\lenthr}   \settowidth{\lenthr}{61811M}

\begin{table}
\begin{center}
{\small\begin{tabular}{llcrrrrc}
\hline\hline
\mcc{OP}     & \mcc{Start Time}      &  \mcc{Length}    & \multicolumn{4}{c}{\Iave\ (C/s)}                      & Orbital \\ \cline{4-7}
\mcc{\#}     & \mcc{(UT)}            &  \mcc{(sec)}     & \mcc{LECS}       & \mcc{MECS}       & \mcc{HPGSPC}    & \mcc{PDS}       & Phase$^a$ \\
\hline\hline
3275         & 07/1/98 14:46:40  &  96582     & $1.331\pm 0.007$ & $3.871\pm 0.010$ & $19.78\pm 0.07$ & $16.24\pm 0.05$ & 0.5837--0.6106 \\
3373         & 17/1/98 22:18:43  &  85375     & $5.581\pm 0.016$ & $14.54\pm 0.020$ & $100.1\pm 0.10$ & $70.34\pm 0.07$ & 0.8322--0.8560 \\
3428         & 23/1/98 13:02:57  &  126695    & $0.397\pm 0.003$ & $2.724\pm 0.007$ & $61.89\pm 0.15$ & $36.24\pm 0.05$ & 0.9674--0.0027 \\
\begin{minipage}{\lenone}3503\\ 3514\end{minipage} &
\begin{minipage}{\lentwo}30/1/98 07:19:12\\ 30/1/98 18:29:37\end{minipage} &
\begin{minipage}{\lenthr}
$\!\left. \begin{array}{c} 15211\\ 61811\end{array}\!\! \right\}$
\end{minipage} &
$0.249\pm 0.004$ & $0.668\pm 0.004$ & $7.394\pm 0.08$ & $6.273\pm 0.05$ & 0.1303--0.1346 \\
3588         & 05/2/98 17:23:39  &  98063     & $0.503\pm 0.005$ & $1.312\pm 0.005$ & $11.94\pm 0.07$ & $10.54\pm 0.05$ & 0.2850--0.3123 \\
3650         & 12/2/98 19:50:07  &  60749     & $1.777\pm 0.012$ & $4.692\pm 0.014$ & $27.43\pm 0.09$ & $23.09\pm 0.06$ & 0.4561--0.4730 \\
\hline
\multicolumn{8}{l}{$^a$ Ephemeris from Sato {\em et~al.} (1986) \nocite{816}}
\end{tabular}}
\caption[]{Log of the observations of \G\ performed by \B\/. Observations 3503
and 3514 have been summed together.}
\label{log}
\end{center}
\end{table}

\subsec{2.1\quad \underline{Spectral Analysis}}

\begin{figure}
\epsfxsize=0.9\textwidth
\epsffile[20 20 570 650]{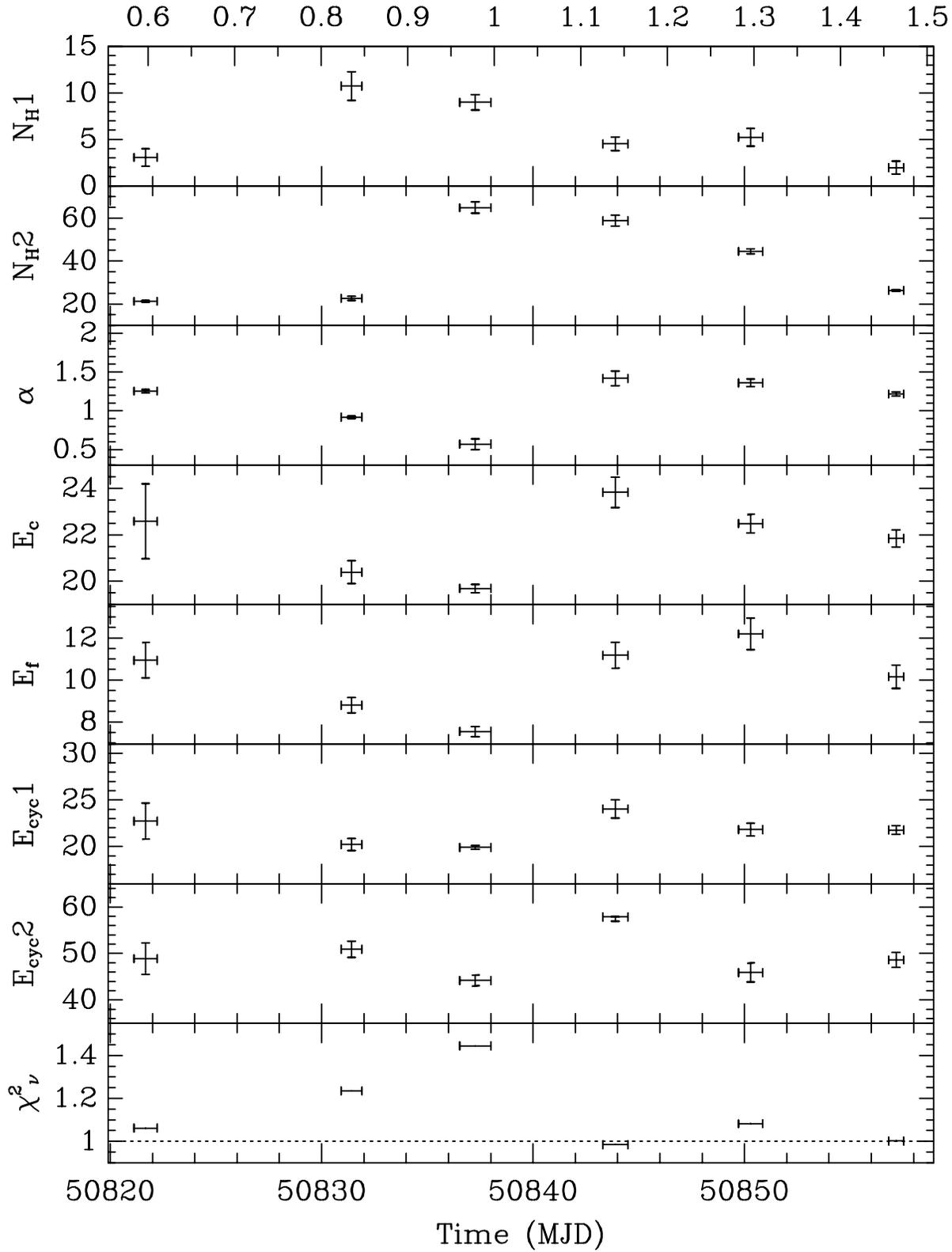}
\vspace{-0.5cm}
\caption[]{Best-fit parameters obtained by fitting two absorbed power laws with
the same index (scattering model in Saraswat et~al.\ (1996)), modified by
a high energy cutoff, plus a gaussian iron line and 2 cyclotron resonance
features. $N_H1$ and $N_H2$ are in units of $10^{22}\ {\rm cm}^{-2}$, while the
energies are in keV. The upper scale refers to the orbital phase from the
ephemeris given by Sato {\em et~al.} (1996).}
\label{fits}
\end{figure}

The spectral analysis of the \B\ observations is very preliminary (we did not
use HPGSPC data). The pulse averaged spectrum is very complex and rich in
features. We show in Fig.~\ref{fits} some of the best-fit parameters  relative
to the fit to the pulse averaged spectra of all the six observations with the
{\em scattering} model by Saraswat {\em et~al.} (1996), defined in terms of
two absorbed power laws, with the same index but different absorptions and
normalizations, plus a high energy cutoff for describing PDS data.
The stronger absorption affects the pulsating, hard (above 4
keV) component, while the softer absorption is though to arise because of
scattering. We added an Iron line at $\sim 6.4$~keV, but we found systematic
deviations at $\sim 20$ and $\sim 40$ keV. While the latter might be
attributable to a cyclotron resonance, the correspondence of the cyclotron
energy with the cutoff energy for the former makes, at this stage of
the analysis an identification with a cyclotron resonance difficult.

\subsec{2.2\quad \underline{Timing Analysis}}

\begin{figure}
\epsfxsize=0.9\textwidth
\epsffile[20 20 570 650]{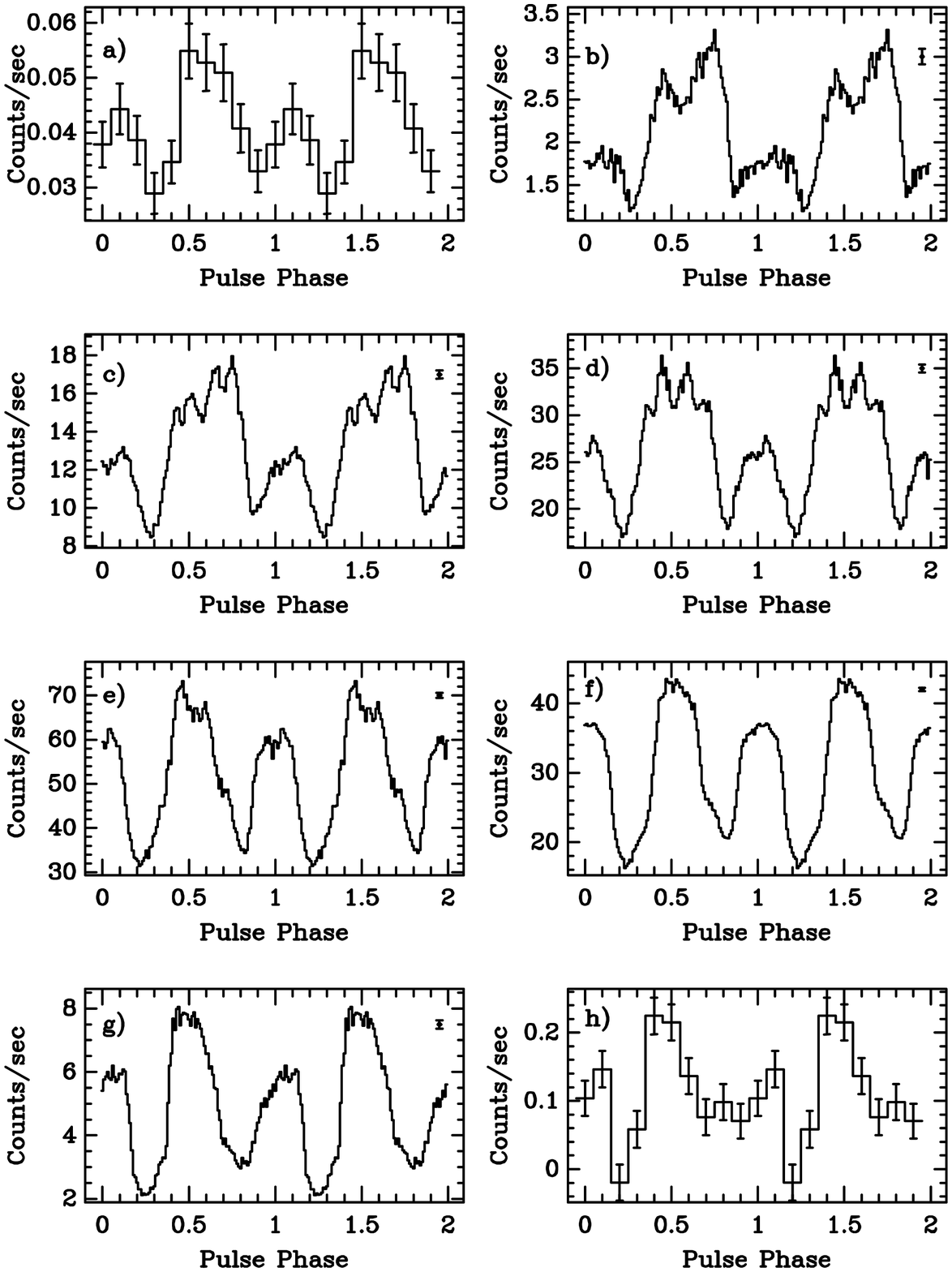}
\vspace{-0.5cm}
\caption[]{\G\ background subtracted pulse profiles as a function of energy for
OP3373, the most intense among all the observations: a) 0.2--2 keV (LECS); b)
1.5--4 keV (MECS); c) 4--10 keV (MECS); d)
4--10 keV (HPGSPC); e) 10--20 keV (HPGSPC); f) 15--30 keV (PDS); g) 30--60 keV
(PDS); h) 60--100 keV (PDS).
}
\label{puls}
\end{figure}

We folded the light curves of all the observations with the apparent pulse
period obtained by an epoch folding search. In this preliminary analysis
arrival times were not corrected to the solar system barycenter nor for orbital
motion. The background subtracted pulse profiles as a function of energy for
the most intense observation are shown in Fig.~\ref{puls}. In this particular
observation we were also able to detect pulsation below 2 keV and in the
60--100 keV range. Note the evolution of the four sub-peaks in the main peak.
The modulation index, defined as $1 - I_{\rm min}/I_{\rm max}$, where $I_{\rm
min}$ and $I_{\rm max}$ are the minimum and maximum count rate observed in the
pulse profile, shows a monotonic increase with energy without any deviation at
the suspected cyclotron resonance features \cite{1286}.

\sec{3\quad REFERENCES}

\end{document}